\documentclass[preprint,nofootinbib,amsmath,prd,aps,superscriptaddress,tightenlines,12pt]{revtex4}
\usepackage{graphicx}
\usepackage{graphics}
\usepackage{amsmath}

\begin{document}
\setlength\baselineskip{17pt}

\begin{flushright}
\vbox{
\begin{tabular}{l}
ANL-HEP-PR-12-xx
\end{tabular}
}
\end{flushright}
\vspace{0.1cm}


\title{\bf Single-variable asymmetries for measuring the `Higgs' boson spin and CP properties}

\vspace*{1cm}

\author{Radja Boughezal}
\email[]{rboughezal@hep.anl.gov}
\affiliation{High Energy Physics Division, Argonne National Laboratory, Argonne, IL 60439, USA} 
\author{Thomas J. LeCompte}
\email[]{lecompte@anl.gov}
\affiliation{High Energy Physics Division, Argonne National Laboratory, Argonne, IL 60439, USA} 
\author{Frank Petriello}
\email[]{f-petriello@northwestern.edu}
\affiliation{High Energy Physics Division, Argonne National Laboratory, Argonne, IL 60439, USA} 
\affiliation{Department of Physics \& Astronomy, Northwestern University, Evanston, IL 60208, USA}


  \vspace*{0.3cm}

\begin{abstract}
  \vspace{0.5cm}
  
  We introduce a class of asymmetries sensitive to the spin and CP properties of the new boson discovered by the ATLAS 
  and CMS experiments.  These asymmetries can be measured in the four-lepton final state, and are defined by integrating the invariant masses of the lepton pairs 
  over specified ranges.  We outline a program of measurements using initial LHC data to determine the 
  quantum numbers and coupling structure, provide analytic expressions for decay widths in several representative models, and discuss what 
  can be determined using the available data.  As examples, we show how the combination of ATLAS and CMS data already disfavor
 certain spin-2 couplings, and discuss how further data will allow for discrimination of a pure CP-odd scalar from a CP-even 
 hypothesis.
  
\end{abstract}

\maketitle


\section{Introduction}
\label{sec:intro}

The recent discovery of a new boson by the ATLAS and CMS collaborations at the LHC~\cite{:2012gk,:2012gu} has ushered in a new era in particle physics.  The future program of the LHC, and the next stage of experimental studies in high energy physics, will be largely devoted to measuring and understanding the properties of the new state in order to determine the 
underlying theory from which it arises.  The initial data provides only a hazy glimpse at the properties of the new particle.  
Observation of its decay into two photons indicates that it cannot be a spin-one state, according to the Landau-Yang theorem~\cite{landau-yang}.  Initial measurements of its branching ratios into various final states indicates that its couplings are consistent with those predicted for the Standard-Model Higgs boson, as determined by the experimental 
collaborations and by several independent analyses~\cite{Carmi:2012in,Plehn:2012iz,Djouadi:2012rh}.  The slight excess observed over Standard Model predictions in the $\gamma\gamma$ final state has already received explanations both within~\cite{Baglio:2012et} and beyond~\cite{Carena:2012xa,Joglekar:2012hb,Giudice:2012pf,Delgado:2012sm,Hashimoto:2012qe,Howe:2012xe,Li:2012jf,Sato:2012bf,Kang:2012bv,Bae:2012am,Davoudiasl:2012ig} the Standard Model.  Significant work will clearly be needed to sharpen our picture of the new state.

Two of the first quantities requiring determinations are the spin and CP properties of the observed particle.  These can 
be measured through a variety of ways in multiple final states.  An initial attempt to determine the CP properties assuming a spin-zero state has been made~\cite{Coleppa:2012eh}, and a discussion of how to disentangle the various spin and CP possibilities once more data is taken was presented in Ref.~\cite{kirill}.  As the largest experimental excesses are in the $\gamma\gamma$ and $ZZ \to 4l$ final states, where either one or both $Z$-bosons are off-shell, initial studies will focus on these two modes.  Two broad categories of techniques exist for the measurement of the new particle's quantum numbers.  Multi-variate 
methods input all of the kinematic information in an event into a likelihood function that can be used to exclude hypotheses 
for the new state's identity.  These ideas have been suggested in the literature for both analyzing the new particle's properties and assisting in discovery~\cite{Gao:2010qx,DeRujula:2010ys,Gainer:2011xz}, and in particular are heavily used in the CMS analysis of the $ZZ$ final state~\cite{:2012gu}.  In the large-time and large-data set limit, such techniques provide the most sensitivity to 
particle properties, since no kinematic information is neglected.  Alternatively, single variables that provide sensitivity to properties of interest can be studied.  Analyses of this type offer the advantages of simplicity and clarity over multi-variate approaches, and can be more easily implemented to provide answers quickly.  They can also indicate which input variables should be used to improve the efficiency of multi-variate techniques.

We introduce here a powerful single-variable measurement in the four-lepton final state that discriminates among both the 
spin and CP possibilities for the new state.  It can be used when either one or both intermediate $Z$ bosons are off-shell, and is relatively insensitive to background contamination.  The idea is simple to explain.  Let $M_{12}$ and $M_{34}$ respectively denote the same-flavor lepton pairs with the highest and lowest invariant masses.  We suggestively call the mass of the heavy resonance decaying to the leptons $M_H$.  $M_{34}$ must satisfy the following inequality: $M_{34} \leq M_H-M_{12}$.  The fall-off of the $M_{34}$ distribution as this upper limit is approached is sensitive to the spin and CP nature of the heavy resonance.  Denoting the 
momentum of the 34-pair in the $H$ rest frame as $\beta$, for a pure CP-even spin-zero state this distribution 
decreases linearly in $\beta$.  For a pure CP-odd spin-zero state, it falls off as $\beta^3$.  For spin-two states, it falls off as either $\beta$, $\beta^3$, or $\beta^5$, depending on the couplings of this state to spin-one particles.  Measurement of this 
distribution provides a powerful handle on the couplings of this new state.  

The sensitivity of the $M_{34}$ distribution to the 
identity of resonances decaying to a pair of $Z$-bosons has been discussed before in the literature~\cite{Choi:2002jk}.  We sharpen and extend this 
observation by defining a class of asymmetries that provide strong discrimination against various hypotheses for spin, CP and 
couplings.  These ``high-low" asymmetries are defined as follows:
\begin{equation}
{\cal A}_{M_{cut}} = \frac{N(M_{34}>M_{cut})-N(M_{34}<M_{cut})}{N(M_{34}>M_{cut})+N(M_{34}<M_{cut})}.
\label{asymdef}
\end{equation}
For a given $M_{12}$, $M_{34} \in [M_{low},M_H-M_{12}]$, where $M_{low}$ is determined by the experimental cuts.  $N$ denotes the number of events in the indicated range of 
$M_{34}$, and $M_{cut}$ can be chosen to provide discrimination between various hypotheses for the $\beta^n$ fall-off.  The measurement of ${\cal A}$ is advantageous for several reasons.  The variables $M_{12}$ and $M_{34}$ are exactly those used in the discovery of the new state, making it easy to perform this analysis quickly with the existing data.  The background in this channel is fairly low.  With very few events the allowed spin and CP combinations are already reduced by measuring ${\cal A}$.  In this paper we investigate this idea in the following ways
\begin{itemize}

\item We provide the decay widths differential in $M_{12}$ and $M_{34}$ in several reference models for use in experimental studies.

\item We outline a program of analyses that can exclude various spin, CP and coupling hypotheses.

\item We demonstrate using the public ATLAS and CMS data that  several spin-2 
coupling possibilities are already disfavored.

\end{itemize}
A word of caution on the last item is necessary.  Even in an an analysis this simple, there are details, subtleties and systematic effects in interpreting the data that can only be properly treated by the experimental collaborations.  
Conclusive evidence for or exclusion of a hypothesis can only be provided by ATLAS and CMS.  Our goal here is only 
to demonstrate the power of the ${\cal A}$ analysis using the available data as an example.  We encourage the 
experimental collaborations to perform a thorough and conclusive study.

Our paper is organized as follows.  We present the differential decay widths for several representative spin and CP combinations in Section~\ref{sec:widths}.  We discuss the construction of the asymmetry in Section~\ref{sec:analysis}, and show how the 
current ATLAS data can be used to discriminate among the various hypotheses.  We discuss our results and outline a program of experimental analysis in Section~\ref{sec:conc}.


\section{Derivation of the decay widths}
\label{sec:widths}

We provide here for completeness a derivation of the decay widths for the heavy-resonance decay into four leptons 
for several representative examples.  We study the mode $H(p_H) \to Z(p_{ee})Z(p_{\mu\mu}) \to e^+(p_{e+}) e^-(p_{e-}) \mu^+(p_{\mu +}) \mu^-(p_{\mu -})$, and extend the result to the case of four same-flavor leptons at the end.  The intermediate $Z$ bosons may be either on-shell or off-shell.  We begin by considering the spin-zero hypothesis for $H$.  The $HZZ$ Feynman rule can be written in the generic form $i T^{\mu\nu}(p_{ee},p_{\mu\mu})$.  It is straightforward to derive the following 
expression for the differential decay width:
\begin{eqnarray}
\frac{d^2 \Gamma}{ds_{ee} \, ds_{\mu\mu}} &=& \frac{G_F M_Z^2}{36\pi^4 M_H} \, \text{BW}(s_{ee}) \, \text{BW}(s_{\mu\mu})
	\int \text{dPS}_2 (p_{ee},p_{\mu\mu}) \, T_{\mu\nu}(p_{ee},p_{\mu\mu})
	T_{\rho\sigma}^{\dagger}(p_{ee},p_{\mu\mu}) \nonumber \\ &\times & s_{ee} s_{\mu\mu} \left\{ g^{\mu\rho}-\frac{p_{ee}^{\mu}p_{ee}^{\rho}}{s_{ee}}\right\}
	\left\{ g^{\nu\sigma}-\frac{p_{\mu\mu}^{\nu}p_{\mu\mu}^{\sigma}}{s_{\mu\mu}}\right\}.
\end{eqnarray}
We have set $s_{ee} = p_{ee}^2 = (p_{e+}+p_{e-})^2$ and $s_{\mu\mu} = p_{\mu\mu}^2 = (p_{\mu+}+p_{\mu-})^2$.  These quantities become the $M_{12}$ and $M_{34}$ introduced in Section~\ref{sec:intro} when a hierarchy between them is 
specified.  We have introduced the following notation for the Breit-Wigner distribution of the $Z$-bosons and the two-particle 
phase space:
\begin{eqnarray}
\text{BW}(s) &=& \frac{1}{(s-M_Z^2)^2+\Gamma_Z^2 M_Z^2}, \nonumber \\
\text{dPS}_2 (p_{ee},p_{\mu\mu}) &=& \int d^4 p_{ee} d^4 p_{\mu\mu}  \delta(p_{ee}^2-s_{ee}) 
	\delta(p_{\mu\mu}^2-s_{\mu\mu}) \, \delta^{(4)}(p_H - p_{ee}-p_{\mu\mu}),
\end{eqnarray}
where $\Gamma_Z$ denotes the $Z$-boson width.

To proceed further, we must specify the form of $T^{\mu\nu}$.  We will consider the two cases of a pure CP-even scalar and a pure CP-odd scalar.  It is possible to also consider a mixed state, but we do not pursue that option here.  The forms for the 
interaction vertices in each case are given below:
\begin{eqnarray}
0^{++} &:& T^{\mu\nu}(p_{ee},p_{\mu\mu}) = a_1 M_Z g^{\mu\nu}, \nonumber \\
0^{-+} &:& T^{\mu\nu}(p_{ee},p_{\mu\mu}) = a_2 \epsilon^{\mu\nu\rho\sigma} p_{ee,\rho} p_{\mu\mu,\sigma}/\Lambda.
\end{eqnarray}
$\Lambda$ has dimensions of energy, while $a_1$ and $a_2$ are dimensionless constants.  We note that for a Standard-Model Higgs boson at tree-level, $a_1 = g/c_{W}$ and $a_2=0$.  Since we will later utilize a shape analysis, the values of these 
parameters do not matter.  Using these expressions it is straightforward to derive the following differential decay widths:
\begin{eqnarray}
\frac{d^2 \Gamma(0^{++})}{ds_{ee} \, ds_{\mu\mu}} &=& |a_1|^2 \frac{G_F M_Z^4}{72\pi^3 M_H} \, \text{BW}(s_{ee}) \, 	\text{BW}(s_{\mu\mu}) \, \beta \left\{ 3s_{ee}s_{\mu\mu}+\frac{\beta^2 M_H^4}{4} \right\}, \nonumber \\
\frac{d^2 \Gamma(0^{-+})}{ds_{ee} \, ds_{\mu\mu}} &=& |a_2|^2 \frac{G_F M_Z^2 M_H^3}{144\pi^3 \Lambda^2} \, \text{BW}(s_{ee}) \, \text{BW}(s_{\mu\mu}) \, \beta^3 s_{ee} s_{\mu\mu}.
\end{eqnarray}
We have introduced the relative momentum of the off-shell $Z^{*}$ bosons in the $H$ rest frame, in units of $M_H$:
\begin{equation}
\beta = \frac{\sqrt{M_H^4+s_{ee}^2+s_{\mu\mu}^2-2 M_H^2 s_{ee}-2 M_H^2 s_{\mu\mu}-2 s_{ee} s_{\mu\mu}}}{M_H^2}.
\end{equation}
We note that $\beta =0$ when $\sqrt{s_{ee}} = M_H-\sqrt{s_{\mu\mu}}$.  As claimed earlier, the $0^{++}$ distribution decreases linearly as this upper limit is approached, while the $0^{-+}$ distribution falls off as $\beta^3$.

We next derive the decay width for a spin-two state.  We write its coupling to $Z$-bosons as $i T^{\mu\nu,\rho\sigma}(p_{ee},p_{\mu\mu})$.  It is straightforward to derive the following differential decay width:
\begin{eqnarray}
\frac{d^2 \Gamma}{ds_{ee} \, ds_{\mu\mu}} &=& \frac{G_F M_Z^2}{36\pi^4 M_H} \, \text{BW}(s_{ee}) \, \text{BW}(s_{\mu\mu})
	\int \text{dPS}(p_{ee},p_{\mu\mu}) \, T_{\alpha\beta,\mu\nu}(p_{ee},p_{\mu\mu})
	T_{\gamma\delta,\rho\sigma}^{\dagger}(p_{ee},p_{\mu\mu}) \nonumber \\  &\times& s_{ee} s_{\mu\mu} 
	\left\{ g^{\mu\rho}-\frac{p_{ee}^{\mu}p_{ee}^{\rho}}{s_{ee}}\right\}
	\left\{ g^{\nu\sigma}-\frac{p_{\mu\mu}^{\nu}p_{\mu\mu}^{\sigma}}{s_{\mu\mu}}\right\} B^{\alpha\beta,\gamma\delta}.
\end{eqnarray}
$B$ denotes the numerator for a massive spin-two particle propagator.  It can be obtained from Refs.~\cite{Giudice:1998ck,Han:1998sg}.  We must again specify a form of the spin-two particle coupling to $Z$-bosons.  A natural choice is to consider a 
spin-two Kaluza-Klein graviton, which we study first.  Another approach is to introduce a model-independent 
parameterization of the coupling, as was followed in Ref.~\cite{Gao:2010qx}.  We will also examine that possibility.

We denote the spin-two graviton with the notation $2^{G}$.  The Feynman rules for this state can be obtained from 
Ref.~\cite{Han:1998sg}.  The differential decay width is given by the following expression:
\begin{equation}
\frac{d^2 \Gamma(2^G)}{ds_{ee} \, ds_{\mu\mu}} = \frac{G_F M_Z^2 \kappa^2}{288\pi^3 M_H} \, \text{BW}(s_{ee}) \, 		\text{BW}(s_{\mu\mu}) \, s_{ee} s_{\mu\mu} \, \beta \, {\cal G}(M_H,M_Z,s_{ee},s_{\mu\mu}).
\end{equation}
$\kappa$ is a constant with dimensions of inverse mass.  Its numerical value will not be important in our analysis.  The 
full expression for the function ${\cal G}$ is lengthy and not especially illuminating.  We present it for completeness in an Appendix.  However, we note here that it does not vanish in the $\beta \to 0$ limit.  We demonstrate this by setting $\sqrt{s_{\mu\mu}}=M_H-\sqrt{s_{ee}}$, which is equivalent to setting $\beta=0$.  We find
\begin{equation}
\frac{d^2 \Gamma(2^G)}{ds_{ee} \, ds_{\mu\mu}} \overbrace{\to}^{\beta \to 0} \frac{G_F M_Z^2 \kappa^2}{288\pi^3 M_H} \, \text{BW}(s_{ee}) \, \text{BW}(s_{\mu\mu}) \, s_{ee} s_{\mu\mu} \, \beta \, \left\{ 40 \left( M_H \sqrt{s_{ee}}-s_{ee}+M_Z^2\right)^2\right\}.
\end{equation}
The spin-two graviton decay width therefore decreases linearly with $\beta$ as $\beta \to 0$.

A spin-two resonance does not necessarily have to be a Kaluza-Klein graviton.  It could instead arise as a composite from a 
strongly-coupled theory, and have different couplings to gauge bosons than the graviton.  A useful parameterization of 
spin-two couplings to vector bosons was presented in Ref.~\cite{Gao:2010qx}.  The generic vertex has ten couplings.  Of these ten couplings, only three lead to a linear decrease in the $M_{34}$ distribution as $\beta \to 0$.  The others lead to distributions that fall off as either $\beta^3$ or 
$\beta^5$ as $\beta \to 0$.  This behavior is simple to understand; the coupling that produces a $\beta^3$ behavior is 
that of a CP-odd ``pseudo-tensor", while those which produce $\beta^5$ behavior have additional momentum insertions and 
therefore are produced in a higher partial-wave state.  As we have already introduced an example with $\beta^3$ dependence 
above, we choose an effective operator which leads to a $\beta^5$ behavior for demonstration purposes:
\begin{equation}
T^{\mu\nu,\rho\sigma}(p_{ee},p_{\mu\mu}) = \frac{1}{\Lambda^3}(p_{ee}-p_{\mu\mu})^{\mu}(p_{ee}-p_{\mu\mu})^{\nu} 
	\left\{p_{ee} \cdot p_{\mu\mu}g^{\rho\sigma} - p_{ee}^{\sigma} p_{\mu\mu}^{\rho} \right\}.
\end{equation}
This corresponds to the coupling $g_4^{(2)}$ in Ref.~\cite{Gao:2010qx}.  We denote this spin-two possibility as $2^{A}$.  It is straightforward to derive the differential decay width of this state:
\begin{equation}
\frac{d^2 \Gamma(2^A)}{ds_{ee} \, ds_{\mu\mu}} = \frac{G_F M_Z^2 M_H^3}{108\pi^3 \Lambda^6} \, \text{BW}(s_{ee}) \, \text{BW}(s_{\mu\mu}) \, s_{ee} s_{\mu\mu}\, \beta^5 \, \left\{ 6 s_{ee} s_{\mu\mu} +\beta^2 M_H^4\right\}.
\end{equation}
This features a much sharper decrease as $\beta \to 0$.

We now comment on the extension to 
the $eeee$ and $\mu\mu\mu\mu$ final states.  These contain both an additional symmetry factor of $1/4$ from the identical particles in the final state, and also arise from two Feynman diagrams rather than a single one as for the $ee\mu\mu$ mode.  As we are interested in a 
shape analysis, the symmetry factor does not affect our results.  It can be simply included in the formulae above if needed. Similarly, each of the individual diagrams squared gives 
expressions for the differential decay width identical to those presented above, and simply contribute a multiplicity factor 
of 2, to which our analysis is again not sensitive.  We now assume that one of the intermediate $Z$-bosons is close to its mass-shell.  This can be checked to be true for the majority of the ATLAS data~\cite{:2012gk}. Then, the squared diagrams will have one of the 
Breit-Wigner functions becoming large: $\text{BW} \sim 1/\Gamma^2$.  The interference term is not enhanced 
in this kinematic region, and can be neglected.  Therefore, the differential decay width presented above can be safely used to perform 
shape analyses in all leptonic final states.  It is simple to extend the formulae above to account for the interference if 
necessary.  

Before proceeding, we note that this analysis can be extended to higher spin states than spin-two.  As discussed in Ref.~\cite{Choi:2002jk}, these fall off quickly as $\beta^{2J-3}$, where $J$ denotes the spin of the state.  We also note that 
we have not discussed the spin correlations between production and decay that occur for the spin-two state.  Since we are only measuring invariant mass distributions of final-state particles, these will not affect our study.


\section{Analysis}
\label{sec:analysis}

\begin{figure}[t]
\centerline{
\includegraphics[height=15.0cm,angle=90]{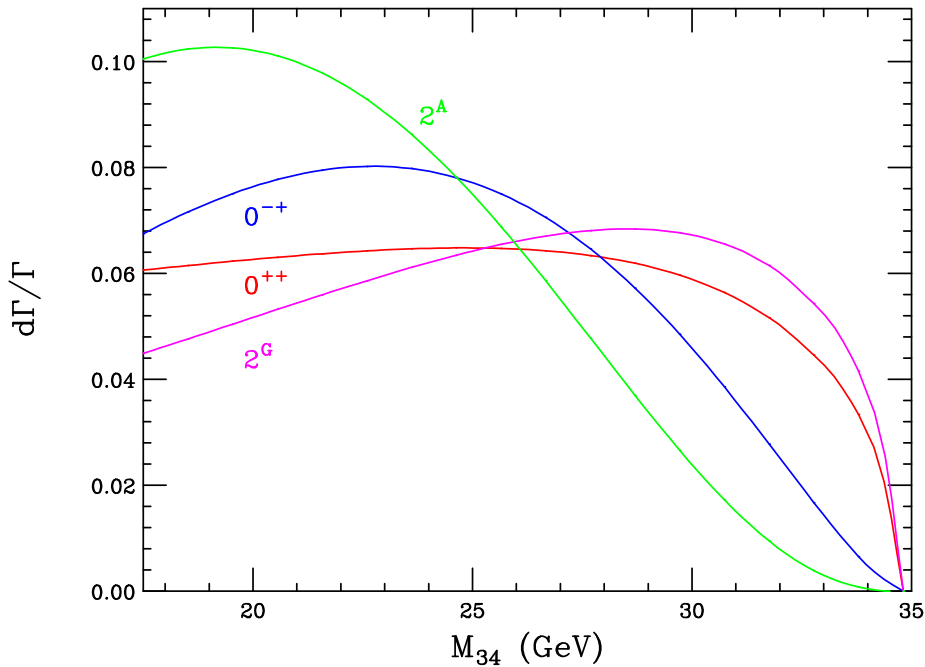}
}
\vspace{-0.5cm}
\caption{The normalized decay width $1/\Gamma_{tot} \times d^2 \Gamma /(dM_{12}^2 dM_{34}^2)$  as a function of $M_{34}$ for the spin-0 CP-even and CP-odd states $0^{++}$ and $0^{-+}$, the spin-2 graviton $2^{G}$, and the alternate spin-two state $2^A$.  The parameter choices are described in the text.}
\label{widthplot}
\end{figure}

We begin our discussion of how to test spin and CP hypotheses for the new boson by presenting the normalized 
differential decay widths in Fig.~\ref{widthplot}.  We have set $M_H=126$ GeV, and for simplicity have set $M_{12}=M_Z$.  The lower invariant mass $M_{34}$ is shown in the range $M_{34} \in [17.5\, \text{GeV},M_H-M_Z]$, where the upper limit corresponds to the kinematically allowed maximum, and the lower limit to that set by ATLAS.  CMS sets their lower limit to 12~GeV.  The normalization of the decay widths is  performed by dividing the expressions presented in the previous section by their integrals over the allowed range in $M_{34}$.  The shape difference induced by the differing $\beta^n$ behavior of the various states is clear in the plot.  The $0^{-+}$ and $2^A$ states are peaked toward lower $M_{34}$, the $0^{++}$ state is flat, and the $2^G$ is peaked toward high $M_{34}$.

It is now simple to construct a single variable capable of discriminating between various hypotheses.  We begin by considering the discrimination between the $0^{++}$ and $2^A$ states.  The two distributions cross at an invariant mass 
$M_{34} \approx 26$ GeV.  It is intuitively clear that we should compare the integrals of the two distributions above and 
below the crossing point.  We form the asymmetry defined in Eq.~(\ref{asymdef}) with $M_{cut} = 26$ GeV.  Choosing 
$M_{cut}$ to be the crossover point can be shown to maximize the sensitivity to the different hypotheses. We compute this for the two states under consideration and find the following results:
\begin{equation}
{\cal A}^{sig}_{26}(0^{++}) = -0.078, \;\;\;
{\cal A}^{sig}_{26}(2^A) = -0.57.
\label{disc:2a:s}
\end{equation}

This result for the signal asymmetry must be combined with the background asymmetry in order to obtain a prediction that 
can be compared to the result measured by the experimental collaborations.  To compute the background asymmetry, we 
first note that the composition of signal and background in the kinematic region of interest is expected to be roughly 1:1 from 
experimental studies~\cite{:2012gk,:2012gu}.  We further note that it is dominated by continuum production of $ZZ^{*}$.  We estimate the background asymmetry by running a leading-order Madgraph~\cite{Alwall:2011uj} simulation of the $pp \to Zl^+ l^-$ process with basic acceptance cuts reproducing those in the ATLAS analysis.  We find an asymmetry of 
${\cal A}^{back} \approx -0.045$.  Assuming that 50\% of the measured rate comes from background events, we find the following 
predictions for the combined signal plus background asymmetries:
\begin{equation}
{\cal A}^{sig+back}_{26}(0^{++}) = -0.060, \;\;\;
{\cal A}^{sig+back}_{26}(2^A) = -0.31.
\label{disc:2a:sb}
\end{equation}

We can now use the available LHC data to estimate the experimental value of ${\cal A}_{26}$ together with the statistical error on the measurement.  We keep events that are $\pm 5$ GeV from the $Z$-peak in order to use the simple choice of $M_{cut}=26$ GeV; we discuss this point further below.  We caution that this is an estimate 
only.  Corrections for detector acceptance, experimental resolutions on the invariant masses, systematic errors and other effects must be considered.  Similarly, events with $M_{12} \neq M_Z$ should be included to increase the sensitivity to different hypotheses.  For simplicity we neglect these events, and comment on extending the analysis to include them later in the text.  Only the experimental collaborations can 
authoritatively perform this analysis.  With these caveats mentioned, ATLAS reports 4 events in the higher mass bin and 4 in the 
lower mass bin; CMS reports 1 and 1. for a combined raw asymmetry of 0.0 with a statistical uncertainty of approximately 0.28.  The $2^A$ hypothesis is disfavored by the data at about the one standard-deviation level.  We note that other $\beta^5$ couplings lead to similar results; for example, 
using the effective coupling corresponding to $g_7^{(2)}$ in Ref.~\cite{Gao:2010qx} leads to an asymmetry 
${\cal A}^{sig}_{26}(g_7^{(2)}) = -0.69$, in even more disagreement with the LHC results.  The message we wish to convey is that decay widths behaving as $\beta^n$ with $n \geq 5$ are unlikely to be compatible with the current data.  

We proceed to consider whether the state $0^{-+}$ is consistent with the current data.  We form the asymmetry ${\cal A}_{28}$, which provides the strongest discrimination between the $0^{++}$ and $0^{-+}$ states.  We find the following 
results using the expressions in Section~\ref{sec:widths}:
\begin{equation}
{\cal A}^{sig}_{28}(0^{++}) = -0.33, \;\;\;
{\cal A}^{sig}_{28}(0^{-+}) = -0.58.
\label{disc:0-+s}
\end{equation}
Treating the background as before, these become 
\begin{equation}
{\cal A}^{sig+back}_{28}(0^{++}) = -0.31, \;\;\;
{\cal A}^{sig+back}_{28}(0^{-+}) = -0.44.
\label{disc:0-+sb}
\end{equation}
The combined experimental asymmetry from ATLAS and CMS is $-0.40\pm 0.27$.  At this point, both possibilities are 
consistent with the data.

Finally, we compare the $2^G$ distribution with the $0^{++}$ hypothesis.  The crossover point for these two distributions is 
roughly 25 GeV.  We compute the ${\cal A}_{25}$ asymmetries for both states and find the following results:
\begin{equation}
{\cal A}^{sig}_{25}(0^{++}) = 0.05, \;\;\;
{\cal A}^{sig}_{25}(2^{G}) = 0.18.
\label{disc:2Gs}
\end{equation}
Treating the background as before, these become
\begin{equation}
{\cal A}^{sig+back}_{25}(0^{++}) = 0.06, \;\;\;
{\cal A}^{sig+back}_{25}(2^{G}) = 0.13.
\label{disc:2Gsb}
\end{equation}
The uncertainty on the experimental result is $\pm 0.29$, about four times the separation between the two hypotheses, so there is currently no discrimination possible between the two hypotheses.

In a more complete analysis, events with $M_{12} \neq M_Z$ would be used.  This would double the available amount of data.  For $M_{12}<M_Z$, the available phase space for $M_{34}$ increases, indicating that the crossover point also changes.  A full analysis should consider $M_{cut}$ as a function of $M_{12}$.  Also, we have selected $M_{cut}$ based on the properties of the signal cross section only.  It is possible that the chosen values could shift slightly if the combined signal and background prediction were used to select 
$M_{cut}$.  We leave these extensions to a more complete experimental analysis utilizing a larger dataset.


\section{Discussion and conclusions}
\label{sec:conc}

We have demonstrated that a simple asymmetry which can be easily formed from the available LHC data provides 
discrimination between the possible spin and CP hypotheses for the new boson recently discovered by ATLAS 
and CMS.  The idea behind the measurement is simple: as the lesser lepton-pair invariant mass reaches its upper kinematic 
limit in the four-lepton final state, the decay width falls off with a characteristic power of $\beta$.  The ``high-low" asymmetry defined in Eq.~(\ref{asymdef}) captures this behavior in a single variable.  We have shown that even the few events measured in the four-lepton final state are beginning to restrict the possible couplings of a spin-two state.  An additional benefit to this approach is that it facilitates a simple combination of data from multiple experiments.

These observations suggest the following program of analysis for the experimental collaborations.  
\begin{itemize}

\item The experimental collaborations should confirm our expectation that the possible spin-two couplings cannot be 
composed solely of those containing a $\beta^5$ dependence.  We have demonstrated this by studying 
two examples from the general parameterization presented in Ref.~\cite{Gao:2010qx}.  A more thorough analysis of the 
possible coupling structures should be performed.

\item We have focused on events with $M_{12} \approx M_Z$ for simplicity of presentation.  A more complete analysis would use all of the available data, and would allow $M_{cut}$ to vary for events with different $M_{12}$.  This would increase 
the sensitivity of the method, and should be investigated. 

\item The currently available data cannot yet test the CP-odd spin-zero hypothesis.  This 
analysis should be pursued as the experimental collaborations observe additional events in the four-lepton channel.

\item It is not possible with the current data to distinguish between the spin-two graviton and the $0^{++}$ hypotheses.  As 
more data is collected, this analysis should be revisited.  Angular distributions will also help discriminate between these hypotheses.

\item With more data it will become possible to extend this analysis to determine whether the observed state has admixtures of the various coupling structures; for example, the collaborations could consider whether a spin-zero state possesses both CP-even and CP-odd couplings.  

\end{itemize}
In summary, we are excited about the possibility of our proposed simple measurement of the properties of the new particle 
discovered at the ATLAS and CMS experiments.  We encourage the experimental collaborations to analyze their data 
using the techniques proposed here.

\bigskip
\noindent
{\large {\bf Acknowledgements}}
\medskip

\noindent
We thank C. Zachos for helpful discussions.  This work is supported by the U.S. Department of Energy, Division of High Energy Physics, under contract DE-AC02-06CH11357 and the grants DE-FG02-95ER40896 and DE-FG02-08ER4153.


\bigskip
\noindent
{\large {\bf APPENDIX}}
\bigskip

\noindent
We present here the function ${\cal G}$ that appears in the differential decay width for a spin-two graviton to $Z$-bosons:

\begin{eqnarray}
{\cal G} &=& \left\{
	\frac{1}{3}\,{\frac { \left( {{\it M_Z}}^{4}+12\,{\it s_{ee}}\,{\it s_{\mu\mu}} \right) {{\it M_H}}^
	{4}}{{\it s_{ee}}\,{\it s_{\mu\mu}}}} \right. \nonumber \\
&+&\frac{2}{3}\,{\frac { \left( 3\,{{\it M_Z}}^{4}{\it s_{\mu\mu}}-9\,{
	\it s_{ee}}\,{{\it s_{\mu\mu}}}^{2}+3\,{{\it M_Z}}^{4}{\it s_{ee}}+40\,{\it s_{ee}}\,{\it s_{\mu\mu}}\,{
	{\it M_Z}}^{2}-9\,{{\it s_{ee}}}^{2}{\it s_{\mu\mu}} \right) {{\it M_H}}^{2}}{{\it s_{ee}}\,{
	\it s_{\mu\mu}}}} \nonumber \\
&-&\frac{2}{3}\,{\frac {-42\,{\it s_{ee}}\,{\it s_{\mu\mu}}\,{{\it M_Z}}^{4}-{\it s_{ee}}\,{{
	\it s_{\mu\mu}}}^{3}+7\,{{\it M_Z}}^{4}{{\it s_{\mu\mu}}}^{2}-{{\it s_{ee}}}^{3}{\it s_{\mu\mu}}+7\,{{
	\it M_Z}}^{4}{{\it s_{ee}}}^{2}}{s_{ee}s_{\mu\mu}}} \nonumber \\
&-&\frac{2}{3}\frac{-34\,{{\it s_{ee}}}^{2}{{\it s_{\mu\mu}}}^{2}+20\,{{\it s_{ee}}}^{
	2}{\it s_{\mu\mu}}\,{{\it M_Z}}^{2}+20\,{\it s_{ee}}\,{{\it s_{\mu\mu}}}^{2}{{\it M_Z}}^{2}}{{\it 
	s_{ee}}\,{\it s_{\mu\mu}}} \nonumber \\
&+&\frac{2}{3}\,{\frac { \left( {\it s_{ee}}-{\it s_{\mu\mu}} \right) ^{2} \left( {
	{\it s_{ee}}}^{2}{\it s_{\mu\mu}}+3\,{{\it M_Z}}^{4}{\it s_{ee}}+{\it s_{ee}}\,{{\it s_{\mu\mu}}}^{2}-20
	\,{\it s_{ee}}\,{\it s_{\mu\mu}}\,{{\it M_Z}}^{2}+3\,{{\it M_Z}}^{4}{\it s_{\mu\mu}} \right) }{{
	\it s_{ee}}\,{\it s_{\mu\mu}}\,{{\it M_H}}^{2}}}\nonumber \\
&+&\left. \frac{1}{3}\,{\frac { \left( {\it s_{ee}}-{\it s_{\mu\mu}}
	 \right) ^{4} \left( 2\,{\it s_{ee}}\,{\it s_{\mu\mu}}+{{\it M_Z}}^{4} \right) }{{\it s_{ee}}
	\,{\it s_{\mu\mu}}\,{{\it M_H}}^{4}}}  \right\}.
\end{eqnarray}	


\end{document}